\documentclass{ws-mpla}

\renewcommand{\draftnote}{ }
\renewcommand{\trimmarks}{ }
\begin{document}

\markboth{D. Ebert, R. N. Faustov \& V. O. Galkin}
{Two-Photon Decay Rates of Heavy Quarkonia}

%
%

\title{\vspace{-1.5cm}
    \begin{flushright}
      \textmd{ HU-EP-03/03}
    \end{flushright}\vspace{1cm}
TWO-PHOTON DECAY RATES OF HEAVY QUARKONIA\\ IN THE RELATIVISTIC
  QUARK MODEL} 

\author{D. EBERT}

\address{Institut f\"ur Physik, Humboldt--Universit\"at zu Berlin,
  Invalidenstr.110, D-10115 Berlin, Germany\\
  debert{@}physik.hu-berlin.de} 

\author{R. N. FAUSTOV$^*$ and V. O. GALKIN$^\dag$}

\address{Institut f\"ur Physik,
  Humboldt--Universit\"at zu Berlin, Invalidenstr.110, D-10115 Berlin,
  Germany\\ and\\
Russian Academy of Sciences, Scientific Council for
Cybernetics, Vavilov Str. 40, Moscow 117333, Russia\\
$^*$faustov@theory.sinp.msu.ru,  $^\dag$galkin@physik.hu-berlin.de}

\maketitle


\begin{abstract}
Two-photon decay rates of pseudoscalar, scalar and tensor states of
charmonium and bottomonium are calculated in the framework of the
relativistic quark model. Both relativistic effects and one-loop
radiative corrections are taken into account. The obtained
results are compared with other theoretical predictions and available
experimental data.  

\end{abstract}

\ccode{PACS Nos.: 13.20.Gd, 13.40.Hq, 12.39.Ki}

\vspace*{12pt}
\noindent
The investigation of the two-photon decay rates of heavy quarkonia
is important for understanding the heavy quark dynamics in
mesons. These decays are sensitive probes of the quarkonium wave
function. In the nonrelativistic limit their rates are proportional to
the square of the wave function or its derivative at the
origin.  However, relativistic effects turn out to be
substantial, especially for charmonium, and significantly modify this
dependence. In this note we consider the two-photon decays in the
framework of the relativistic quark model.   

\begin{figure}[th]
\centerline{\psfig{file=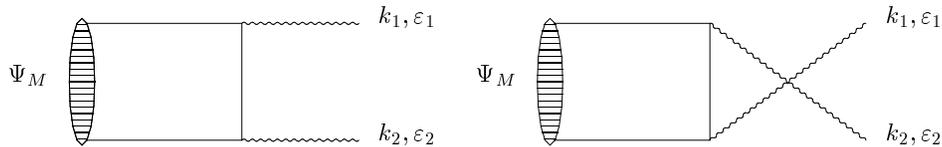,width=12.5cm}}
\vspace*{8pt}
\caption{Two-photon annihilation diagrams of the quarkonium.}
\end{figure}

The diagrams of the $Q\bar Q$ bound state with mass $M$ ($Q=c,b$)
annihilation into 
two photons with momenta $k_1$, $k_2$ and polarization vectors
$\varepsilon_1$, $\varepsilon_2$ are graphically presented in Fig.~1.
The corresponding amplitude in the rest frame of the decaying meson is then
given  by 
\begin{eqnarray}
  \label{eq:me2g}
{\cal M}_{\gamma\gamma}&=&4\pi\alpha\sqrt{3}e_Q^2\int\!\!\frac{d^3p}
{(2\pi)^3}
\sum_{s,t}\bar v^s_Q(-{\bf p})\Biggl[\not\!\varepsilon_2\frac{\not\! p-
\not\! k_1+m_Q}
{(p-k_1)^2-m_Q^2}\not\! \varepsilon_1\cr
&& +(1\leftrightarrow 2)\Biggr]
u^t_Q({\bf p}) \Psi^{s,t}_M({\bf p}),
\end{eqnarray}
where $\alpha=e^2/(4\pi)$ is  the fine structure constant; $e_Q$,
$m_Q$ and ${\bf p}$ are the charge (in units of the electron charge
$e$), mass and relative momentum of quarks; 
$\sqrt{3}$ is  the colour factor and the Dirac spinors of quark and
antiquark are defined by
\begin{eqnarray}
\label{spinor}
&&u^t_Q({\bf p})=\sqrt{\frac{\epsilon_Q(p)+m_Q}{2\epsilon_Q(p)}}
\left(
\begin{array}{c}1\cr {\displaystyle\frac{\bm{\sigma}
      {\bf  p}}{\epsilon_Q(p)+m_Q}}
\end{array}\right)\chi^t,\cr\nonumber\\
&&v^s_Q(-{\bf p})=\sqrt{\frac{\epsilon_Q(p)+m_Q}{2\epsilon_Q(p)}}
\left(
\begin{array}{c} {-\displaystyle\frac{\bm{\sigma}
      {\bf  p}}{\epsilon_Q(p)+m_Q}}\cr 1
\end{array}\right)(-i\sigma_2)\chi^s,
\end{eqnarray}
with  $p^0=\epsilon_Q(p)=\sqrt{{\bf p}^2+m_Q^2}$. The photon momenta in
the chosen frame obey the relations: ${\bf k}_1=-{\bf k}_2$ and 
$k_1^0=k_2^0=|{\bf k}_1|=|{\bf k}_2|=M/2$.    
The meson wave function for the state 
with the orbital momentum $L$, spin $S$ and total momentum $J$ has the
form 
\begin{equation}
  \label{eq:wf}
\Psi_M^{s,t}({\bf p})=\sum_{m,\lambda}\left<LSm\lambda|JM_J\right>
\left<{\textstyle\frac12}{\textstyle\frac12}st|S\lambda\right>
Y_{L\, m}(\theta,\varphi)\phi_M(p),
\end{equation}
where $\phi_M(p)\equiv \phi_M(|{\bf p}|)$ is the radial wave function.

For the calculation of the heavy quarkonium wave functions we use the
relativistic quark model based on the quasipotential approach in
quantum field theory. In this model,\cite{efg,efg1,egf} a meson is
described by the wave 
function of the bound quark-antiquark state, which satisfies the
quasipotential equation of the Schr\"odinger type in
the center-of-mass frame:
\begin{equation}
\label{quas}
{\left(\frac{b^2(M)}{2\mu_{R}}-\frac{{\bf
p}^2}{2\mu_{R}}\right)\Psi_{M}({\bf p})} =\int\frac{d^3 q}{(2\pi)^3}
 V({\bf p,q};M)\Psi_{M}({\bf q}),
\end{equation}
where the relativistic reduced mass
is
\begin{equation}\label{mur}
\mu_{R}=\frac{M^4-(m^2_1-m^2_2)^2}{4M^3},
\end{equation}
and 
$b^2(M)$  denotes
the on-mass-shell relative momentum squared
\begin{equation}
\label{bm}
{b^2(M) }
=\frac{[M^2-(m_1+m_2)^2][M^2-(m_1-m_2)^2]}{4M^2}.
\end{equation}
Here $m_{1,2}$ and $M$ are quark masses
and a meson mass, respectively. For the $Q\bar Q$ bound system
(quarkonium) $m_1=m_2=m_Q$ and Eqs. (\ref{mur}), (\ref{bm}) take
the form
\begin{equation}
  \label{eq:bb}
  \mu_R=\frac{M}{4}, \qquad b^2(M)=\frac{M^2}{4}-m_Q^2.
\end{equation}

The kernel
$V({\bf p,q};M)$ in Eq.~(\ref{quas}) is the quasipotential operator of
the quark-antiquark interaction. It is constructed with the help of the
off-mass-shell scattering amplitude, projected onto the positive
energy states. 
The quasipotential for the heavy quarkonia, including retardation and
one-loop radiative corrections, and expanded in $p^2/m_Q^2$, can be found
in Refs.~\refcite{efg,efg1}. As the initial approximation in the
configuration space we obtain the Cornell potential:
\begin{equation}
  \label{eq:corn}
  V(r)=-\frac43\frac{\bar\alpha_V}r+Ar+B,
\end{equation}
where $\bar\alpha_V$ is the effective coupling constant:
$\bar\alpha_V=0.450$ for charmonium and $\bar\alpha_V=0.285$ for
bottomonium, respectively. In our model the quark masses
$m_b=4.88$ GeV, $m_c=1.55$ GeV, $m_s=0.50$ GeV, $m_{u,d}=0.33$ GeV,
the slope of the linear potential $A=0.18$ GeV$^2$ and the constant
term $B=-0.16$ GeV have fixed values which agree with the usually
accepted ones. In 
the recent paper\cite{efg1} we calculated the mass spectra of heavy
quarkonia with complete account of the relativistic corrections of
order $p^2/m_Q^2$, including retardation effects, and one-loop radiative
corrections. The obtained results agree with experimental data within
few MeV. Now we apply the found relativistic wave functions to
the calculation of the two-photon decay rates of heavy quarkonia.

The two-photon decay amplitudes (\ref{eq:me2g}) of the heavy quarkonium
pseudoscalar (${}^1\!S_0$), scalar (${}^3\!P_0$) and tensor
(${}^3\!P_2$) states have the following structure 
\begin{eqnarray}
  \label{eq:das}
  {\cal M}_{\gamma\gamma}({}^1\!S_0\to \gamma\gamma)&=&g
  \epsilon_{\mu\nu\rho\sigma} k_1^\mu k_2^\nu \varepsilon_1^\rho
  \varepsilon_2^\sigma,\nonumber\\  
 {\cal M}_{\gamma\gamma}({}^3\!P_0\to \gamma\gamma)&=& g_0
  [(\varepsilon_1\cdot \varepsilon_2)](k_1\cdot k_2) -(\varepsilon_1
  \cdot k_2)(\varepsilon_2\cdot k_1)],\nonumber\\ 
 {\cal M}_{\gamma\gamma}({}^3\!P_2\to \gamma\gamma)&=& g_1[(k_1
  \cdot k_2)\varepsilon_1^\mu\varepsilon_2^\nu - (\varepsilon_1\cdot
  k_2) k_1^\mu\varepsilon_2^\nu\nonumber\\ 
&& - (\varepsilon_2\cdot
  k_1) k_2^\mu\varepsilon_1^\nu +(\varepsilon_1\cdot \varepsilon_2)
  k_1^\mu k_2^\nu]\xi^{\mu\nu}\nonumber\\ 
&&+g_2 [(\varepsilon_1\cdot
  \varepsilon_2) (k_1\cdot k_2)-(\varepsilon_1\cdot
  k_2)(\varepsilon_2\cdot  k_1)]k_1^\mu k_2^\nu \xi^{\mu\nu}/
  (k_1\cdot k_2),\qquad
   \end{eqnarray}
where $g_i$ are corresponding decay constants and $\xi^{\mu\nu}$ is
the polarization tensor of the tensor meson. Performing angular
integrations, using the angular dependence of the wave functions given
by Eq.~(\ref{eq:wf}), and summing over photon
polarizations\footnote{For decays of tensor ${}^3\!P_2$ states it is
  also necessary to average over states with different projection $M_J$ of
  the total momentum $J$.}  we get the 
following relativistic expressions for the decay rates\cite{fgq88}: 
\begin{equation}
  \label{eq:2gs0}
  \Gamma^{\rm R}({}^1\!S_0\to
  \gamma\gamma)=\frac{3\alpha^2e_Q^4}{M^2}\Biggl|\int\!\! 
  \frac{d^3p}{(2\pi)^3} \frac{m_Q^2}{p\epsilon_Q(p)}
  \ln\frac{\epsilon_Q(p)+p}{\epsilon_Q(p)-p} \ \phi_S(p)\Biggr|^2,
 \end{equation}

\begin{eqnarray}
  \label{eq:2gp0}
  \Gamma^{\rm R}({}^3\!P_0\to
  \gamma\gamma)&=&\frac{12\alpha^2e_Q^4}{M^2}\Biggl|\int\!\! 
  \frac{d^3p}{(2\pi)^3} \frac{m_Q^2}{Mp}\Biggl[\frac{p}{\epsilon_Q(p)}
  \ln\frac{\epsilon_Q(p)+p}{\epsilon_Q(p)-p}\cr\nonumber\\
&&  +\left(1-\frac{M}{2\epsilon_Q(p)}\right)
\left(2- \frac{\epsilon_Q(p)}{p}
\ln\frac{\epsilon_Q(p)+p}{\epsilon_Q(p)-p}\right)\Biggr]\phi_P(p)\Biggr|^2,
 \end{eqnarray}

\begin{eqnarray}
  \label{eq:2gp2}
 \Gamma^{\rm R}({}^3\!P_2\to \gamma\gamma)&=&\frac{36\alpha^2e_Q^4}{5M^2}
\Biggl\{\Biggl|\int\!\!
  \frac{d^3p}{(2\pi)^3} \frac{m_Q\epsilon_Q(p)}{Mp}\Biggl[\Biggl(2+
\frac{p}{\epsilon_Q(p)}
  \ln\frac{\epsilon_Q(p)+p}{\epsilon_Q(p)-p}\cr \nonumber\\&& 
-\frac{\epsilon_Q(p)}{p}\ln\frac{\epsilon_Q(p)+p}{\epsilon_Q(p)-p}
\Biggr)
\left(1+\frac{m_Q}{2\epsilon_Q(p)[\epsilon_Q(p)+m_Q]}\right)
\cr \nonumber\\&& -\frac{2p^2}
{3\epsilon_Q(p)[\epsilon_Q(p)+m_Q]}\Biggr] \phi_P(p)\Biggr|^2 
\cr \nonumber\\&&+\frac16 \Biggl|\int\!\!
  \frac{d^3p}{(2\pi)^3} \frac{m_Q^2}{Mp}\Biggl[\frac{p}{\epsilon_Q(p)}
  \ln\frac{\epsilon_Q(p)+p}{\epsilon_Q(p)-p} 
+3\left(2-\frac{\epsilon_Q(p)}{p}\ln\frac{\epsilon_Q(p)+p}
{\epsilon_Q(p)-p}\right)\cr \nonumber\\&&-\frac{\epsilon_Q(p)}{m_Q}
\left(2-\frac{M}{m_Q}
\right)\left(2-\frac{\epsilon_Q(p)}{p}\ln\frac{\epsilon_Q(p)+p}
{\epsilon_Q(p)-p}\right)\cr\nonumber\\
&&
+\frac{p^2}{m_Q[\epsilon_Q(p)+m_Q]}\left(2-\frac{M}{m_Q}\right)
\Biggl(\frac12\left(\frac{\epsilon_Q(p)}{p}\ln\frac{\epsilon_Q(p)+p}
{\epsilon_Q(p)-p}-2\right)\cr\nonumber\\
&&\times\left(1-\frac{3\epsilon_Q^2(p)}{p^2}\right)
+1\Biggr)\Biggr]\phi_P(p)\Biggr|^2\Biggr\}.
  \end{eqnarray}
Similar relativistic expressions have also been found in
Refs.~\refcite{bst,lcb,gjr}.

In the nonrelativistic limit $p/m_Q\to 0$ and $M\to 2m_Q$ these
expressions reduce to the known relations\cite{kmrr}
\begin{eqnarray}
  \label{eq:2gnr}
\Gamma^{\rm NR}({}^1\!S_0\to \gamma\gamma)&=&\frac{3\alpha^2e_Q^4
\left|R_{nS}(0)\right|^2}{m_Q^2},\\\label{eq:2gnr0}
\Gamma^{\rm NR}({}^3\!P_0\to \gamma\gamma)&=&\frac{27\alpha^2e_Q^4
\left|R'_{nP}(0)\right|^2}{m_Q^4},\\\label{eq:2gnr2}
\Gamma^{\rm NR}({}^3\!P_2\to \gamma\gamma)&=&\frac{36\alpha^2e_Q^4
\left|R'_{nP}(0)\right|^2}{5m_Q^4}.
\end{eqnarray}
The first order QCD radiative corrections to the two-photon decay
rates can be accounted for by multiplying the relativistic decay rates
$\Gamma^{\rm R}$ (\ref{eq:2gs0})--(\ref{eq:2gp2}) by
the corresponding factors\cite{kmrr}   
\begin{eqnarray}
  \label{eq:radcorr}
 \Gamma({}^1\!S_0\to \gamma\gamma)&=&
\Gamma^{\rm R}({}^1\!S_0\to \gamma\gamma)\left[1+\frac{\alpha_s}{\pi}
\left(\frac{\pi^2}3-\frac{20}3\right)\right],\\\label{eq:radcorr0}
\Gamma({}^3\!P_0\to \gamma\gamma)&=&
\Gamma^{\rm R}({}^3\!P_0\to \gamma\gamma)\left[1+\frac{\alpha_s}{\pi}
\left(\frac{\pi^2}3-\frac{28}9\right)\right],\\\label{eq:radcorr2}
\Gamma({}^3\!P_2\to \gamma\gamma)&=&
\Gamma^{\rm R}({}^3\!P_2\to \gamma\gamma)\left[1
  -\frac{16}3\frac{\alpha_s}{\pi} \right].
\end{eqnarray}
For the numerical analysis we use the two-loop expression for the QCD
coupling constant $\alpha_s$ with $\Lambda_{\overline{\rm
MS}}^{(4)}=175$~MeV obtained\cite{kmrr} from the experimental ratio of  
$\Gamma(\Upsilon\to gg\gamma)/\Gamma(\Upsilon\to ggg)$. This yields
$\alpha_s(m_b)=0.18$ for bottomonium and $\alpha_s(m_c)=0.26$ for
charmonium, respectively. 

\begin{table}[htbp]
\tbl{Two-photon decay rates of heavy quarkonium pseudoscalar
  (${}^1\!S_0$), scalar (${}^3\!P_0$) and tensor (${}^3\!P_2$) states
  (in keV). Comparison of different theoretical predictions and
  experimental data.}
{\begin{tabular}{@{}c@{\ \ \ }c@{\ \ \ }c@{\ \ \ }c@{\ \ \ }c@{\ \ \
      }c@{\ \ \ }c@{}c@{\ \ \ }c@{\ \ \ }c@{\ \ }c@{}}
\toprule
Particle & \multicolumn{6}{c}{Theory}&&\multicolumn{3}{c}{Experiment}\\
\cline{2-7}  \cline{9-11}\\
& our&Ref.\refcite{munz}& Ref.\refcite{gjr}&Ref.\refcite{sbg}
&Ref.\refcite{hlt}&Ref.\refcite{ab} &
&PDG\cite{pdg} &E835\cite{e835}&Belle\cite{belle}\\ 
\colrule
$\eta_c(1{}^1\!S_0)$& 5.5 & 3.5 &10.94& 7.8& 5.5 &4.8 & &$7.5(8)$
&$4.4(1.7)(1.5)$& \\
$\eta'_c(2{}^1\!S_0)$&1.8& 1.38 & & 3.5 & 2.1 & 3.7 & & & &\\
$\chi_{c0}(1{}^3\!P_0)$&2.9&1.39&6.38&2.5&5.32 & & &$3.1(8)$
&$2.90(59)(66)(30)$& \\
$\chi'_{c0}(2{}^3\!P_0)$&1.9&1.11& & & & & & & &\\
$\chi_{c2}(1{}^3\!P_2)$&0.50&0.44&0.57&0.28&0.44& & &$0.45(8)$
&$0.31(5)(4)$&$0.61(10)$\\
$\chi'_{c2}(2{}^3\!P_2)$&0.52&0.48& & & & & & &\\
\colrule
$\eta_b(1{}^1\!S_0)$&0.35&0.22&0.46&0.46&0.45&0.17& & & &\\
$\eta'_b(2{}^1\!S_0)$&0.15& 0.11& &0.20 &0.21&0.13 & & & &\\
$\eta''_b(3{}^1\!S_0)$&0.10&0.084& & & & & & &\\
$\chi_{b0}(1{}^3\!P_0)$&0.038&0.024&0.080&0.043& & & & & &\\
$\chi'_{b0}(2{}^3\!P_0)$&0.029&0.026& & & & & & & &\\
$\chi_{b2}(1{}^3\!P_2)$&0.008&0.0056& 0.008& 0.0074& & & & & &\\
$\chi'_{b2}(2{}^3\!P_2)$&0.006&0.0068& & & & & & & &\\
\botrule
\end{tabular}}
\end{table}

In Table~1 we compare our predictions for two-photon decay rates of
heavy quarkonia calculated with the account of both relativistic
and radiative corrections  with previous theoretical
calculations and experimental averages from PDG\cite{pdg} and very
recent measurements by E835\cite{e835} and Belle\cite{belle}
Collaborations. 
The application of the quark-hadron duality to 
sum rules\cite{ger} for two-photon decays yielded the following
predictions  $\Gamma(\eta_c\to\gamma\gamma)\cong
7.4$~keV, $\Gamma(\chi_{c0}\to\gamma\gamma)\le 6.1$~keV,
$\Gamma(\chi_{c2}\to\gamma\gamma)\le 3.9$~keV. 

It follows from Eqs.~(\ref{eq:2gnr0}), (\ref{eq:2gnr2}) that the ratio
of two-photon decay rates of scalar and tensor states in the 
nonrelativistic approximation is independent of meson wave functions
$${\cal R}_{Q\bar Q}^{\rm NR}\equiv\frac{\Gamma^{\rm NR}({}^3\!P_0\to
\gamma\gamma)}
{\Gamma^{\rm NR}({}^3\!P_2\to\gamma\gamma)} =\frac{15}4=3.75.$$
Inclusion of radiative corrections (\ref{eq:radcorr0}),
(\ref{eq:radcorr2}) considerably increases this ratio
$$ R_{Q\bar Q}^{\rm NR}= \frac{15}4 \frac{\displaystyle \left[
  1+\frac{\alpha_s}{\pi} \left(\frac{\pi^2}3-\frac{28}9\right)\right]}{  
    \displaystyle \left[1-\frac{16}3\frac{\alpha_s}{\pi}\right]}.$$
Indeed for the chosen values of $\alpha_s$ we find  $ R_{c\bar
  c}^{\rm NR}\cong 6.8$ and  $ R_{b\bar b}^{\rm NR}\cong 5.5$. On the other
hand, relativistic effects tend to decrease this ratio. The results of
our calculations of the ratio $R_{Q\bar Q}\equiv
  \Gamma({}^3\!P_0\to\gamma\gamma)/\Gamma({}^3\!P_2\to\gamma\gamma)$
  using Eqs.~(\ref{eq:2gp0}), (\ref{eq:2gp2}), (\ref{eq:radcorr0}),
(\ref{eq:radcorr2}) are given in Table~2 in comparison with other
  theoretical predictions and experimental data.

\begin{table}[htbp]
\tbl{Ratio $R_{Q \bar Q}$ of two-photon decay rates of scalar
  (${}^3\!P_0$) and tensor 
(${}^3\!P_2$) states. Comparison of different theoretical predictions and
  experimental data.}
{\begin{tabular}{@{}cccccccc@{}}\toprule
 & our &Ref.\refcite{munz}& Ref.\refcite{gjr}&Ref.\refcite{sbg}
&Ref.\refcite{hlt}&Ref.\refcite{mp}& CLEO\cite{cleo}\\
\colrule
$R_{c\bar c}(1P)$ & 5.8 & 3.2 & 11.2 & 8.9 & 12.1 & 5.4(1.3)&
 7.4(2.4)(5)(9)\\
$R_{c\bar c}(2P)$ & 3.7 & 2.3 & & & & &  \\
$R_{b\bar b}(1P)$ & 4.8 & 4.3 & 10 &  5.8 & & & \\
$R_{b\bar b}(2P)$ & 4.6 & 3.8 & & & & &  \\
\botrule
\end{tabular}}
\end{table}

Our analysis shows that both relativistic effects and radiative
corrections play an important role in the investigation of two-photon
annihilation rates of heavy quarkonia. Their inclusion considerably
modifies theoretical predictions and brings them in accord with
experimental data. For computations we use the relativistic
quarkonium wave functions which were previously obtained in
calculating the mass spectra of charmonium and bottomonium. The
consistent account of the relativistic corrections to the decay
amplitudes and the use of the relativistic quarkonium wave functions
calculated  in the framework of the
relativistic quark model, which correctly predicts meson mass spectra
and various decay rates, as well as inclusion of radiative corrections
distinguish our approach from previous calculations. The increase of the
experimental precision in measuring the charmonium two-photon decay
rates and the observation of corresponding bottomonium decays will
help to discriminate between different theoretical models.          

\section*{Acknowledgments}
 The authors express their gratitude to S. Eidelman,   
M. M\"uller-Preussker and V.~Savrin  
for support and discussions. R.N.F. and V.O.G. are also grateful to the
organizers of the {\it  Workshop on Quarkonium}, CERN, 8-10 November
2002, and especially to N.~Brambilla and A.~Vairo, for the invitation
to this very productive meeting, which inspired this work,
and for the members of Quarkonium Working Group for useful
discussions. Two of us (R.N.F. and V.O.G.) 
were supported in part by the {\it Deutsche
Forschungsgemeinschaft} under contract Eb 139/2-2.

\end{document}